\documentstyle[11pt,aaspp4]{article}
\input{psfig}
\eqsecnum
\def\beq{\begin{equation}}
\def\eeq{\end{equation}}
\def\ref{\reference}
\def\simge{\mathrel{%
   \rlap{\raise 0.511ex \hbox{$>$}}{\lower 0.511ex \hbox{$\sim$}}}}
\def\simle{\mathrel{
   \rlap{\raise 0.511ex \hbox{$<$}}{\lower 0.511ex \hbox{$\sim$}}}}

\begin{document}
\title{Constraints on Torque-Reversing Accretion-Powered X-ray Pulsars}
\author{Insu Yi$^1$ and J. Craig Wheeler$^{2,3}$}
\affil{$^1$Institute for Advanced Study, Princeton, NJ 08540; yi@sns.ias.edu}
\affil{$^2$Department of Astronomy, University of Texas, Austin, Texas 78712;
wheel@astro.as.utexas.edu}
\affil{$^3$Institute for Theoretical Physics, University of California,
Santa Barbara, 93106-4030}

\begin{abstract}

The observed abrupt torque reversals in X-ray pulsars, 
4U 1626-67, GX 1+4, and OAO 1657-415, can be explained by transition in 
accretion flow rotation from Keplerian to sub-Keplerian, which takes place
at a critical accretion rate, $\sim 10^{16}-10^{17}g/s$.
When a pulsar system spins up near equilibrium spin before the transition, 
the system goes into spin-down after transition to sub-Keplerian.
If a system is well into the spin-up regime, the transition can cause 
a sharp decrease in spin-up rate but not a sudden spin-down. 
These observable types of abrupt torque change are distinguished from the 
smooth torque variation caused by change of ${\dot M}$ in the Keplerian 
flow. The observed abrupt torque reversal is expected when the pulsar magnetic 
field $B_*\sim 5\times 10^{11}b_p^{-1/2}L_{x,36}^{1/2}P_{*,10}^{1/2}G$
where the magnetic pitch parameter $b_p\sim$ a few, $L_{x,36}$ is the X-ray 
luminosity in $10^{36} erg/s$, and $P_{*,10}$ is the pulsar spin period in 
10s. Observed quasi-periodic oscillation (QPO) periods tightly constrain the
model. For 4U 1626-67, ${\dot M}\approx 2.7\times 10^{16}g/s$ with 
$b_p^{1/2}B_*\approx 2\times 10^{12}G$. We estimate ${\dot M}\sim 6\times
10^{16} g/s$ and $b_p^{1/2}B_*\sim 5\times 10^{13}G$ for GX 1+4, and
${\dot M}\sim 1\times 10^{17} g/s$ and $b_p^{1/2}B_*\sim 2\times 10^{13}G$
for OAO 1657-415. Reliable detection of QPOs before and after torque reversal 
could directly test the model.

\end{abstract}


\section{Introduction}

Sudden torque reversal events in some accretion-powered X-ray pulsars 
such as 4U 1626-67, GX 1+4, and OAO 1657-415 have recently been detected
(e.g. Chakrabarty et al. 1993,1996,1997ab). 
The spin-up and spin-down rates are puzzlingly similar despite abrupt torque
reversal. The torques remain largely steady before and after reversal, 
which plausibly indicates the existence of an ordered, stable accretion flow.
These systems are distinguished from those showing random torque fluctuations 
seen in some wind-fed pulsar systems 
(e.g. Nagase 1989, Anzer \& B{\"o}rner 1995). In the torque-reversing systems, 
the mass accretion rate ${\dot M}$ appears gradually modulated with a small 
amplitude on a time scale $\simge yr$ which is much longer than the typical 
reversal time scale $\simge days$. The recently reported flux and spectral
changes around the time of the reversal in 4U 1626-67 shows that the
${\dot M}$ seems to change by about a few $10\%$ in the X-ray emitting region 
close to the neutron star (Vaughn \& Kitamoto 1997). 

Recently, Yi, Wheeler, \& Vishniac (1997)
\footnote{An erratum is to be published in ApJL due to errors in the quoted
parameters. The corrected parameters include substantially higher magnetic
fields and mass accretion rates which are consistent with those in this
work.}
have suggested an explanation for the
torque reversal phenomenon. The reversal is triggered by transition of the 
accretion flow from Keplerian rotation to sub-Keplerian rotation. 
The transition occurs when ${\dot M}$ crosses the critical rate 
${\dot M}_c\sim 10^{16}-10^{17} g/s$. 
For magnetized pulsar systems, the inner region 
of accretion flow lies roughly at a radius $R_o\sim 5\times 10^8 B_{*,12}^{4/7}
{\dot M}_{16}^{-2/7}cm$, where ${\dot M}_{16}={\dot M}/10^{16}g/s$ 
and $B_{*,12}$ is the stellar field strength in units of $10^{12}G$.
This radius is similar to the size of a white dwarf, which strongly suggests
that the transition could be similar to that of cataclysmic variables
(e.g. Patterson \& Raymond 1985, Narayan \& Popham 1993).
The details of the transition process remain unknown.
The relevant time scale for the transition is likely to be the thermal time 
scale $t_{th}\sim (\alpha\Omega_K)^{-1}$ or the viscous-thermal time scale
$t_{vt}\sim (R/H)t_{th}$, where $\alpha$ is the conventional $\alpha$
viscosity coefficient, $H$ is the thickness of the accretion disk 
at radius $R$ from the star, and $\Omega_K=(GM_*/R^3)^{1/2}$ is the Keplerian 
rotation rate for the stellar mass $M_*$. The time scales $t_{th}<t_{vt}\sim 
10^3s$ for $\alpha\sim 0.3$, $R\sim 10^9cm$, and ${\dot M}_{16}=1$ 
(Frank, King, \& Raine 1992). After the transition, the rotation of the 
accretion flow becomes sub-Keplerian due to large internal pressure
(cf. Narayan \& Yi 1995). The change of rotation induces
an abrupt ($\simle day$) decrease of torque exerted on the star by the accretion
flow. The observed X-ray emission is not expected to be significantly affected
since ${\dot M}$ change is small.

We show how the proposed model tightly constrains the pulsar system parameters.
The estimated parameters are largely consistent with other available
estimates. We also derive a simple criterion which identifies candidate
pulsar systems by their observable parameters. Combining this with an
additional constraint from the quasi-periodic oscillation (QPO), we suggest
a possible test of the model. We also discuss different types of torque
variation and how they differ observationally from each other. The observed 
reversal events are likely to occur in pulsar systems near spin equilibrium 
with ${\dot M}\sim {\dot M}_c\sim$ a few $\times 10^{16} g/s$. We take the 
neutron star moment of inertia $I_*=10^{45} g~cm^2$, radius $R_*=10^{6} cm$,
and $M_*=1.4M_{\odot}$. The magnetic field is assumed to be a dipole 
configuration (Frank et al. 1992). The angular velocity of the star 
$\Omega_*=2\pi/P_*$.

\section{Accretion Flow Transition and Torque Reversal}

When the accretion flow rotation is Keplerian, the corotating stellar 
magnetic field lines interact with the rotating accretion flow with 
$\Omega_{K}(R)$ (e.g. Yi et al. 1997, Wang 1995 and references therein). 
In a steady state, the pitch of the azimuthally stretched magnetic field 
is given by
\beq
{B_{\phi}(R)\over B_z(R)}
={\gamma\over\alpha}{\Omega_*-\Omega_{K}(R)\over\Omega_{K}(R)},
\eeq
where $\gamma\le 1$ measures the vertical velocity shear length scale.
The dimensionless ratio $b_p=\gamma/\alpha$, which determines the magnetic
pitch, is not well constrained, but is likely to be order unity 
(e.g Wang 1995 and reference therein). The magnetically disrupted accretion 
flow gives an inner edge at $R=R_o$ determined by
\beq
\delta=\left(R_o\over R_c\right)^{7/2}
\left[1-\left(R_o\over R_c\right)^{3/2}\right]^{-1}
\eeq
where $R_c=(GM_*P_*^2/4\pi^2)^{1/3}$ is the Keplerian corotation radius and
\beq
\delta={2(2\pi)^{7/3}R_*^5\over
(GM_*)^{2/3}}{\left(\gamma/\alpha\right)B_*^2\over P_*^{7/3}L_x}
\approx 2.1\times 10^{-2}B_{eff,11}^2P_{*,10}^{-7/3}L_{x,36}^{-1}.
\eeq
We have defined $B_{eff,11}=b_p^{1/2}B_*/10^{11}G$ and $L_x=GM_*{\dot M}/R_*$.
The torque exerted on the star is
\beq
N={7\over 6}N_0{1-(8/7)(R_o/R_c)^{3/2}\over 1-(R_o/R_c)^{3/2}}
\eeq
where $N_0={\dot M}(GM_*R_o)^{1/2}$ (Yi et al. 1997, Wang 1995). 
The spin equilibrium $N=0$ is achieved when $R_o/R_c=x_{eq}=(7/8)^{2/3}$.
The pulsar spin evolution follows ${\dot P_*}=-P_*^2N/2\pi I_*$
as ${\dot M}$ varies, which we model as a linear variation determined
simply by $d{\dot M}/dt$.
If the accretion flow remains Keplerian with gradually varying ${\dot M}$,
the disruption radius and the resulting torque $N$ varies according to
eqs. (2-2),(2-4). 
If a pulsar system evolves from a spin-up (spin-down) state and ${\dot M}$ 
decreases (increases) gradually, it is possible that $R_o/R_c=x_{eq}$ is
reached at a certain ${\dot M}$ and then moves to $R_o/R_c>x_{eq}$ ($<x_{eq}$)
and hence spin-down (spin-up). As long as ${\dot M}$ variation is smooth and 
gradual the torque reversal event is expected to be smooth.
Figure 1 shows an example of the smooth transition for 4U 1626-67, which
is clearly distinguished from the observed abrupt reversal.

When the decreasing ${\dot M}$ reaches ${\dot M}_c$ below which the
accretion flow becomes sub-Keplerian (Yi et al. 1997), the accretion 
flow-magnetic field interaction would change on a time scale $\simle day$. 
For sub-Keplerian rotation, $\Omega/\Omega_K=A<1$, the new corotation radius
$R_c^{\prime}=A^{2/3}R_c$ and the new inner edge $R_o^{\prime}$ is determined by
\beq
\delta^{\prime}=\left(R_o^{\prime}\over R_c^{\prime}\right)^{7/2}
\left[1-\left(R_o^{\prime}\over R_c^{\prime}\right)^{3/2}\right]^{-1}
\eeq
where $\delta^{\prime}=\delta A^{-7/3}$.
The new torque on the star after the transition is
\beq
N^{\prime}={7\over 6}N_0^{\prime}
{1-(8/7)(R_o^{\prime}/R_c^{\prime})^{3/2}\over
1-(R_o^{\prime}/R_c^{\prime})^{3/2}},
\eeq
where $N_0^{\prime}=A{\dot M}(GM_*R_o^{\prime})^{1/2}$.
As in the Keplerian flow, 
$N^{\prime}=0$ would occur when $R_o^{\prime}/R_c^{\prime}=x_{eq}$.
In a sub-Keplerian flow supported by internal pressure, $\Omega=A\Omega_K$
is largely determined by the ratio of magnetic to gas pressure
in the accreted plasma. For equipartition between magnetic and gas pressures, 
with the pressure ratio, say, $0.05-1$, we expect $A\approx 0.14-0.4$ for all 
$\alpha=0.01-0.3$ (e.g. Narayan \& Yi 1995). 
For the Keplerian flow, the equilibrium spin period is
$P_{eq}\approx 13(b_p/10)^{3/7}B_{*,12}^{6/7}{\dot M}_{16}^{-3/7}s$. 
For the sub-Keplerian rotation, the new equilibrium spin period 
$P_{eq}^{\prime}=P_{eq}/A>P_{eq}$. A system in a spin-up state with a
restricted period ratio given by
$1\simle P_*/P_{eq}\simle A^{-1}$ before transition, would 
evolve toward $P_{eq}^{\prime}\simge P_*$ after transition, i.e.
a sudden torque reversal. Figure 1 shows a fit to the observed reversal
event in 4U 1626-67 (cf. Yi et al. 1997). Vaughn and Kitamoto (1997)
report that ${\dot M}$ changes by $\sim 20$\% before and after the
the observed reversal. The model in Figure 1 is consistent with this
finding. On the other hand, if a system has a pre-transition 
$P_*/P_{eq}>A^{-1}$ that places it well into the spin-up regime, 
the transition could merely lead to a sudden decrease of spin-up rate,
but not to spin-down 
(i.e. $P_*>P_{eq}^{\prime}$), a behavior most likely in systems with
smaller $B_*$. Figure 1 shows a hypothetical pulsar similar to 4U 1626-67 
except with a smaller $B_*$. Observations of this type of torque change could 
support the present accretion flow transition model.

Quasi-periodic oscillation (QPO), if observed, could provide an additional 
constraint. Such a constraint would depend critically on the nature of the 
QPO mechanism. Adopting the widely used beat frequency model 
(e.g. Lamb et al. 1985), we get the QPO periods
\beq
P_{QPO}=P_*\left[(R_o/R_c)^{-3/2}-1\right]^{-1}, \qquad
P_{QPO}^{\prime}=P_*\left[(R_o^{\prime}/R_c^{\prime})^{-3/2}-1\right]^{-1}
\eeq
before and after transition, respectively. 
It remains unclear whether the conventional beat frequency model can
explain the recently observed kHz QPOs in some X-ray pulsar systems.
The QPO periods we use in the following section are much longer than
than the kHz QPO periods. There exists no indication that the long period
QPOs arise from other physical processes.

\section{Application to Observed X-ray Pulsar Systems}

The expressions derived so far can be readily applied to the observed
systems.

4U 1626-67 has $P_*\approx 7.660s$ at reversal and the observed torques
for the adopted $I_*$ are $N\approx 5.37\times 10^{33} g cm^2/s^2$ and
$N^{\prime}\approx -4.51\times 10^{33} g cm^2/s^2$, or the observed torque
ratio $(N^{\prime}/N)_{obs}\approx -0.840$ 
(Chakrabarty 1996, Chakrabarty et al. 1997a). 
Using the observed $P_{QPO}\approx 25s$ during spin-up (Shinoda et al.
1990), eq. (2-7) gives
$R_o/R_c\approx 0.84$ or $\delta\approx 2.3$ from eq. (2-2). Using the
observed $N$ and the derived $R_o/R_c$ in eq. (2-4), we get 
${\dot M}\approx 2.7\times 10^{16}g/s$. Making use of the observed 
$N^{\prime}$ and 
$\delta^{\prime}=\delta A^{-7/3}$, we solve eqs. (2-5), (2-6) to obtain
$A\approx 0.46$, which suggests that the accreted plasma has an equipartition 
strength magnetic field. From eq. (2-3), we then find 
$B_{eff}=b_p^{1/2}B_*\approx 1.7\times 10^{12}G$. 
This field strength is smaller than the estimates
$B_*\sim (6-8)\times 10^{12}G$ obtained by Pravado et al. (1979) and
Kii et al. (1986) based on the energy cutoff in X-ray spectrum and the energy
dependence of the pulse shape. Although uncertainties in the latter
estimates of $B_*$ are not clear,
a simple comparison suggests that $b_p^{1/2}$ is at most
order unity (cf. Wang 1995). The observed 0.7-60 keV flux $F_x\sim 2.4\times 
10^{-9} erg/s/cm^2$ (Pravado et al. 1979) 
gives the distance estimate $(L_x/4\pi F_x)^{1/2}\sim 4.2$
kpc, which is consistent with the previous estimates 
(e.g. Chakrabarty 1996). Eq. (2-7) suggests a possible $P_{QPO}^{\prime}\sim 
120s$ after transition. Detection of such a QPO could directly test the model.

GX 1+4 reversed torque from spin-down to spin-up around the spin period 
$P_*\approx 122.15s$ (Chakrabarty 1995, Chakrabarty et al. 1997b) with the 
measured torques $N\approx 3.77\times 10^{34} g cm^2/s^2$, and 
$N^{\prime}\approx -3.14\times 10^{34}gcm^2/s^2$, which give 
$(N^{\prime}/N)_{obs}\approx -0.833$.
A QPO of period $P_{QPO}\sim 250s$ was reported during the 1976 spin-up
(Doty et al. 1981), but none near the recent reversal episode.
The applicability of this QPO is therefore questionable. Nevertheless,
assuming that the 1976 spin-up state is similar to the recent spin-up, we
can adopt this QPO to make estimates for GX 1+4. The spin period 
around 1976 is close to the recent spin period, which may help to justify
our use of the 1976 QPO data. Following the same 
procedure as for 4U 1626-67, we estimate that $R_o/R_c\sim 0.77$, 
$\delta\sim 1.2$, and ${\dot M}\sim 2.9\times 10^{16} g/s$. We get
$A\sim 0.12$, which again indicates a magnetic field strength near
equipartition. 
Finally from eq. (2-3), $B_{eff}\sim 3\times 10^{13}G$ or $B_{*}\sim 10^{13}G$ 
for $b_p\sim 10$. The observed flux $F_x\sim 2\times 10^{-10} erg/s/cm^2$
in the range of 20-60 keV (Chakrabarty et al. 1997b) gives a distance estimate 
of $\sim 15kpc$ for the derived ${\dot M}$ whereas for the Doty et al. (1981)
value of $F_x\sim 8\times 10^{-9} erg/s/cm^2$ in the 1.5-55 keV range gives
a distance estimate of $\sim 2.4kpc$. 
During the recent spin-down, $P_{QPO}^{\prime}\sim 6160s$ would be
predicted.
This prediction, however, is much less certain than that for
4U 1626-67 due to the lack of reliable $P_{QPO}$ measurement near
the recent spin-up. We note that the recent GX 1+4 reversal
is considerably more gradual than the 4U 1626-67 event. It is possible that
the smooth transition type shown in Figure 1 could be relevant in this case
(Yi et al. 1997).

OAO 1657-415 has an observed pulse period $P_*\approx 37.665s$ at the time of
reversal. Recent observations give $N\approx 4.40\times 10^{34} g cm^2/s^2$ and 
$N^{\prime}\approx -1.06\times 10^{34}gcm^2/s^2$ (Chakrabarty et al. 1993) or
$(N^{\prime}/N)_{obs}\approx -0.241$. There exists no reported QPO for this 
system. If we take $A=0.4$ based on our estimates in
4U 1626-67 and GX 1+4, the observed torque 
ratio $(N^{\prime}/N)_{obs}$ corresponds to $\delta\sim 1.2$ or 
$R_o/R_c\sim 0.77$ as in GX 1+4, which gives ${\dot M}\sim 1.0
\times 10^{17}g/s$. Therefore, we get $B_{eff}\sim 2\times 10^{13}G$ or 
$B_*\sim 10^{13}G$ for $b_p\sim 10$. For $P_*\sim 37s$, $R_o/R_c\sim 0.77$, 
and $A=0.4$, we expect $P_{QPO}\sim 75s$ and $P_{QPO}^{\prime}\sim 430s$ 
before and after reversal, respectively. For a detected flux of $\sim
10^{-9} erg/s/cm^2$ (White \& Pravado 1979, Mereghetti et al. 1991) 
and the estimated ${\dot M}\sim 1\times 10^{17} g/s$, the distance $\sim 12$kpc.

We caution that the predicted QPO frequencies may not be easily detectable.
During spin-down, the accretion flow thickens considerably and
the pulsed polar emission may be significantly diluted by scattering
(Yi et al. 1997). The long period QPOs may appear as occasional flares with 
repetition time scales of $\sim$a few $10^3s$. 

\section{Necessary Condition for Torque Reversal in Systems Near Spin 
Equilibrium}

The parameters derived above can be understood by a simple expression 
that constrains the pulsar parameters and essentially determines the pulsar 
magnetic field. We assume that the system reverses its torque from spin-up 
to spin-down. Just before reversal, we take $R_o/R_c=x_{eq}-\epsilon$ with 
$\epsilon <x_{eq}$. Expanding eqs. (2-2),(2-4) to first order in 
$\epsilon/x_{eq}$ and combining we get
\beq
N={7N_0\over 4}{\epsilon/x_{eq} \over 1-x_{eq}^{3/2}}=
{7\over 2(7-4x_{eq}^{3/2})}\left[1-{1-x_{eq}^{3/2}\over x_{eq}^{7/2}}\delta
\right].
\eeq
After transition to the sub-Keplerian state, we take
$R_o^{\prime}/R_c^{\prime}=x_{eq}+\epsilon^{\prime}$ also with 
$\epsilon^{\prime}< x_{eq}$.
We expand eqs. (2-5),(2-6) to first order in $\epsilon^{\prime}/x_{eq}$
and combine them to derive
\beq
N^{\prime}=-{7N_0^{\prime}\over 4}{1\over 1-x_{eq}^{3/2}}{\epsilon^{\prime}
\over x_{eq}}=-{7N_0^{\prime}\over 2(7-4x_{eq}^{3/2})}
\left[{1-x_{eq}^{3/2}\over x_{eq}^{7/2}}\delta^{\prime}-1\right].
\eeq
The ratio of the sub-Keplerian spin-down torque ($N^{\prime}$) to the
Keplerian spin-up torque ($N$) is then
\beq
{N^{\prime}\over N}=-A^{4/3}\left[1+{g(x_{eq})(1-x_{eq}^{3/2})\over 
4(7/4-x_{eq}^{3/2})}(\delta^{\prime}-\delta)\right]
{g(x_{eq})\delta^{\prime}-1\over 1-g(x_{eq})\delta}
\eeq
where $g(x_{eq})=(1-x_{eq}^{3/2})/x_{eq}^{7/2}$. We have made use of
the first order expression
\beq
{N_0^{\prime}\over N_0}=A^{4/3}\left(1+{\epsilon^{\prime}\over x_{eq}}
+{\epsilon\over x_{eq}}\right)
=A^{4/3}\left[1+{g(x_{eq})(1-x_{eq}^{3/2})\over 4(7/4-x_{eq}^{3/2})}
(\delta^{\prime}-\delta)\right].
\eeq
For $x_{eq}=(7/8)^{2/3}$, we get
\beq
{N^{\prime}\over N}
=-A^{4/3}\left[1+6.10\times 10^{-3}(\delta^{\prime}-\delta)\right]
{0.171\delta^{\prime}-1\over 1-0.171\delta}.
\eeq

Given the observational constraint $N^{\prime}\simle N$, eq. (4-5) indicates
that small values of $A\sim 0.1$ are unlikely, which is consistent with
the result $A\ge 0.3$ in the previous section. 
This in turn suggests that the magnetic to gas pressure ratio is roughly at
the level of equipartition (cf. Narayan \& Yi 1995).
From eqs. (4-1),(4-2), the first necessary condition for the torque reversal 
is $N^{\prime}/N<0$ or
\beq
A^{7/3}{x_{eq}^{7/2}\over 1-x_{eq}^{3/2}}<\delta<{x_{eq}^{7/2}\over 
1-x_{eq}^{3/2}}
\eeq
or for $x_{eq}=(7/8)^{2/3}$
\beq
\delta_{min}=5.86A^{7/3}<\delta<5.86.
\eeq
The second condition comes from the observational requirement,
$\left|N^{\prime}/N\right|_{obs}<1$, which gives (cf. eq. 4-5)
\beq
\delta<5.86{1+A^{4/3}\over 1+A^{-1}}.
\eeq
Finally, the observed continuous X-ray emission throughout the whole reversal 
process requires the accretion to be continuous. 
This translates into the condition 
$R_o/R_c<1$ and $R_o^{\prime}/R_c^{\prime}<1$ in order to avoid an angular
momentum barrier at the corotation radius. By expressing $\epsilon^{\prime}$
in terms of $\delta$, we get
\beq
\delta<13.39A^{7/3}=\delta_{max}.
\eeq
For $A\le 0.51$, $\delta_{max}$ replaces the upper bounds in eqs. 
(4-7),(4-8). For $A\ge 0.51$, the upper bound on $\delta$ is given by 
eq. (4-8) as shown in Figure 2. As $A\rightarrow 1$, the allowed parameter
space shrinks rapidly, which simply indicates that the rotation gets closer
to the Keplerian and the spin-down becomes more difficult when the mass 
accretion rate is fixed. Therefore, the rotation needs to be substantially 
sub-Keplerian after transition in order for the reversal to occur.
For most of the $A$ values, $\delta$'s range is limited to be
within a factor of $\sim 2$.
Therefore the condition for torque reversal becomes $\delta_{min}<\delta<
\delta_{max}$ and the magnetic field should be close to
\beq
B_{eff}\approx 7\times 10^{11}\delta^{1/2}L_{x,36}^{1/2}P_{*,10}^{7/6}G
\eeq
where $\delta_{min}^{1/2}=2.4A^{7/6}<\delta^{1/2}
<3.7A^{7/6}=\delta_{max}^{1/2}$. If the accreted plasma is magnetized with
near equipartition between magnetic pressure and gas pressure, the sub-Keplerian
flow with $A\sim 0.1-0.4$ (Narayan \& Yi 1995) suggests that $\delta^{1/2}$
is essentially a constant of order unity (cf. Figure 2). 
Therefore, eq. (4-10) provides an 
estimate of the magnetic field within a factor of $\sim 2$. 
If the distance to the pulsar and an independent estimate on $B_*$ (say through
X-ray spectrum and emission) are known, this relation can constrain
$b_p=(B_{eff}/B_{*})^2$. The pulsar parameters derived in the previous section 
are consistent with eq. (4-10). This can be checked as follows. 
Using eqs. (4-5), we derive
\beq
\delta={A^{4/3}\over 0.1707}{1-A^{-4/3}(N^{\prime}/N)_{obs}\over
A^{-1}-(N^{\prime}/N)_{obs}}.
\eeq
For a given $A$, we confirm that the observed torque ratio gives $\delta$'s 
which are close enough to the derived values in the previous section.
Therefore we conclude that eq. (4-10) can be reliably used to identify
the characteristics of the accreting pulsar systems 
that should show torque reversals. 

The condition for the torque reversal, $5.86A^{7/3}<\delta<13.39A^{7/3}$,
for $A\le 0.51$ allows a very tightly constrained physical parameters of 
the pulsar systems. Based on the results in the previous section, our model
suggests $A\approx 0.4$ and therefore $0.69<\delta<1.58$. Then, in eq. (4-10), 
$0.8<\delta^{1/2}<1.3$ is determined accurately. If this is the case,
we can derive $B_{eff}\approx 7\times 10^{11} L_{x,36}^{1/2}P_{*,10}^{7/6}G$
essentially without uncertainties in $\delta$. Since $\delta\propto A^{7/3}$, 
$B_{eff}\propto A^{7/6}$ has a crucial dependence on $A$.

\section{Discussion}

The current analysis and the study of Yi et al. (1997) suggest that 
the observed torque reversals occur when the accretion flow makes a 
transition at a critical rate ${\dot M}_c\sim 10^{16}-10^{17} g/s$. 
This points to an interesting connection between X-ray pulsar systems 
and other compact accretion systems such as cataclysmic variables 
and black hole transients. We speculate that the spectral transitions 
seen in the latter systems may be due to a
common accretion flow transition mechanism (cf. Narayan \& Yi 1995). 
The observed torque reversal could be a signature of a pulsar system near 
spin equilibrium with ${\dot M}$ gradually varying near ${\dot M}_c$. 
The proposed mechanism tightly constrains the pulsar field strength. 
If a pulsar does not have an appropriate field strength (eq. 4-10),
it would not undergo a torque reversal even if ${\dot M}$ passes
through ${\dot M}_c$ causing the transition.
If all systems have a similar ${\dot M}_c$ (Narayan \& Yi 1995), 
the pulsar magnetic field strength can be determined despite a major 
uncertainty in the distance. It is possible that
some systems with $P_*>P_{eq}$ abruptly change their torques without
reversing sign at ${\dot M}\sim {\dot M}_c$. These types of
sudden torque change are distinguishable from the smooth torque 
variation in a Keplerian accretion flow. If QPOs can be detected near
torque reversal, the proposed model could be further tested in detail.

Although the spectral and flux changes are generally small in the
torque reversing systems (Chakrabarty 1996), which indicates the
small changes in ${\dot M}$, at least in 4U 1626-67, some spectral and 
flux changes have been reported (Vaughn \& Kitamoto 1997). The hardening
of the spectrum after the transition to spin-down could be attributed
to the accretion flow transition itself (Yi et al. 1997). It is however
unclear if the hot accretion flow with a large vertical thickness
could directly affect the X-ray spectrum through scattering since
the expected scattering depth through the hot flow is at most a few
$\times 10^{-2}$ for ${\dot M}\sim 2\times 10^{16} g/s$ (Narayan \& Yi
1995) appropriate for 4U 1626-67.

Gradual ${\dot M}$ modulation on a time scale ranging
from $\sim yr$ to a few decades is required for the proposed torque reversal.
Such a modulation can be driven by several known mechanisms.
(i) In the case of GX 1+4, the binary orbital motion on a time scale
$\sim yr$ is plausible (Chakrabarty 1996), which could cause ${\dot M}$
variation through orbital modulation of mass transfer. In 4U 1626-67, 
an orbital time scale longer than a few years is unlikely 
(Rappaport et al. 1977, Joss et al. 1978, Shinoda et al. 1990). 
OAO 1657-415 has an orbital time scale of $\sim 10d$ (Chakrabarty et al. 1993).
(ii) Several precession time scales involving
the pulsar and the secondary remain viable (e.g. Thorne et al. 1986).
(iii) X-ray irradiation-induced mass flow oscillation 
(Meyer \& Meyer-Hofmeister 1990) could provide long time scale ($>month$) 
oscillations when certain conditions such as disk size and $\alpha$ are met.
Interestingly, in 4U 1626-67 the observed optical pulsation frequency is
the same as that in X-rays, which has been attributed to the reprocessing
of X-rays by the accretion disk (Ilovaisky et al. 1978, Chester 1979).
(iv) The disk instability model (e.g. Smak 1984) has not been applied to
neutron star systems and long term ${\dot M}$ variation due to the disk 
instability needs further investigation.

\acknowledgments
We acknowledge receiving useful references from L. Bildsten and D. Chakrabarty.
JCW appreciates the hospitality of the Institute for Theoretical Physics
which is supported by NSF grant No. PHY94-07194. This work has been supported 
in part by SUAM Foundation (IY) and by NSF grant AST95-28110 (JCW).

\vfill\eject

\centerline{Figure Caption}
\vskip 1cm

{\bf Figure 1:}
Left columns: Torque reversal event of 4U 1626-67. The solid line is the
combined BATSE and pre-BATSE spin period history adopted from Chakrabarty
(1996). The dotted lines correspond to the model with the accretion flow 
transition where $B_*=5.4\times 10^{11}$G, the mass accretion rate decreases 
from an initial rate of ${\dot M}_i=3.2\times 10^{16} g/s$ at a rate
$d{\dot M}/dt=-6\times 10^{14}g/s/yr$. The accretion flow makes a transition 
at ${\dot M}_c=2.4\times 10^{16}g/s$. The dashed line corresponds to the 
Keplerian model without accretion flow transition where $B_*=10^{12}$G,
${\dot M}_i=5.9\times 10^{16}g/s$, and $d{\dot M}/dt=-2.35\times 10^{15}g/s/yr$.
The observed event is noticeably more sudden than the latter model. 
Right Column: A hypothetical pulsar similar to 4U 1626-67 (with abrupt reversal)
except $B_*=3\times 10^{11}G$. When $P_*>P_{eq}/A$ (i.e. well into spin-up),
the accretion flow transition results in abrupt decrease of torque without 
reversal (dotted line). The dashed line is the corresponding Keplerian model
without transition. The parameters are $b_p=10$ and $A=0.46$ (see text).

\vskip 1cm

{\bf Figure 2:}
Allowed region in the $\delta-A$ parameter space for the torque reversal.
The region between the two lines are allowed for the torque reversal.

\vfill
\eject

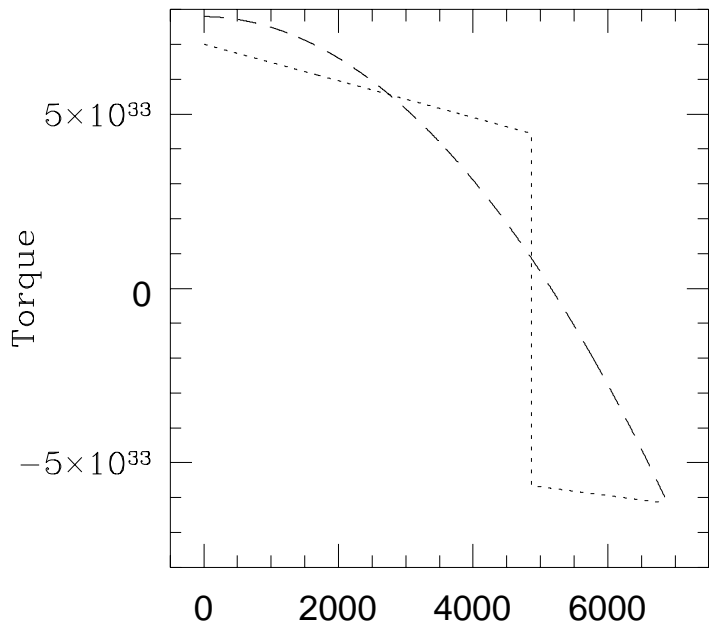
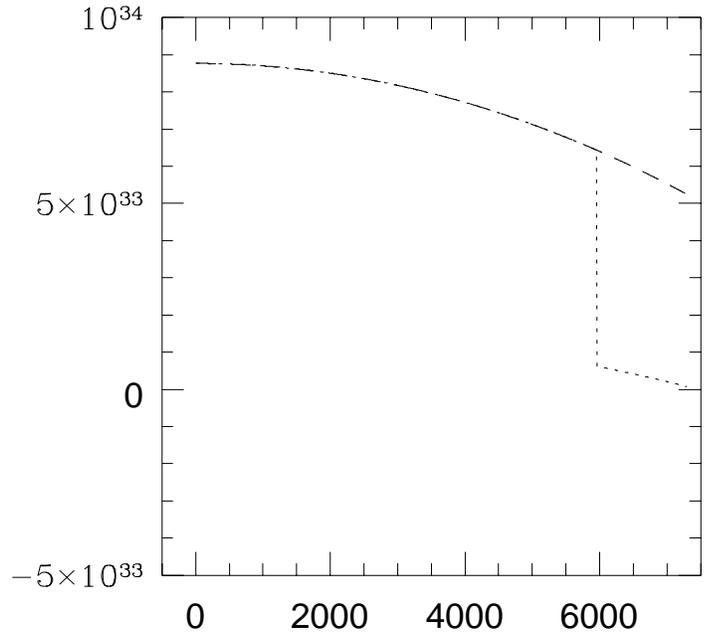
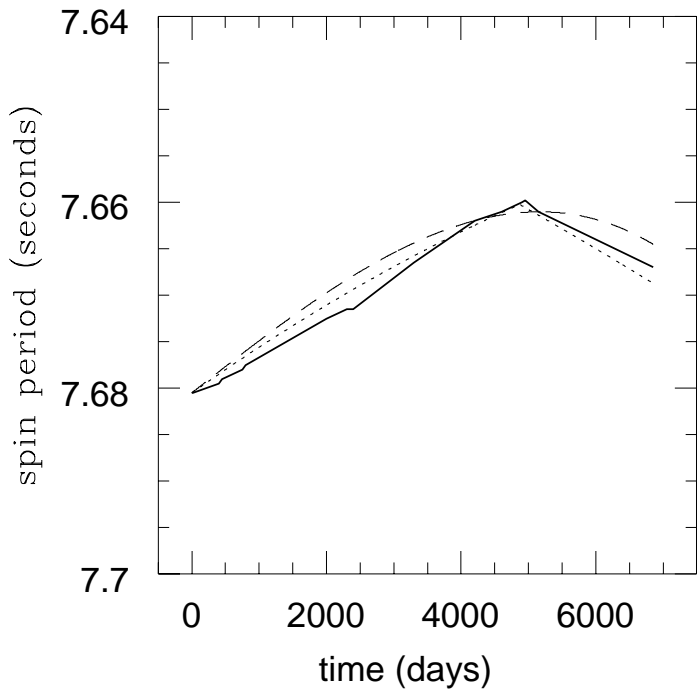
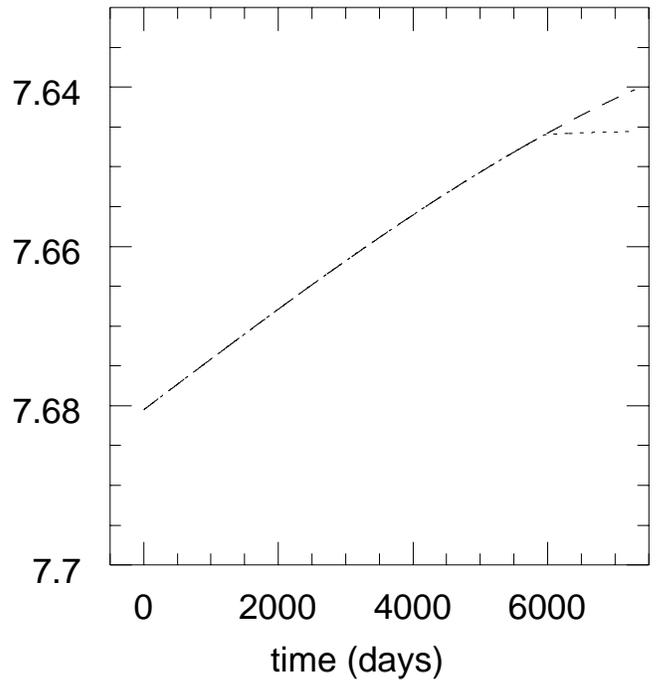
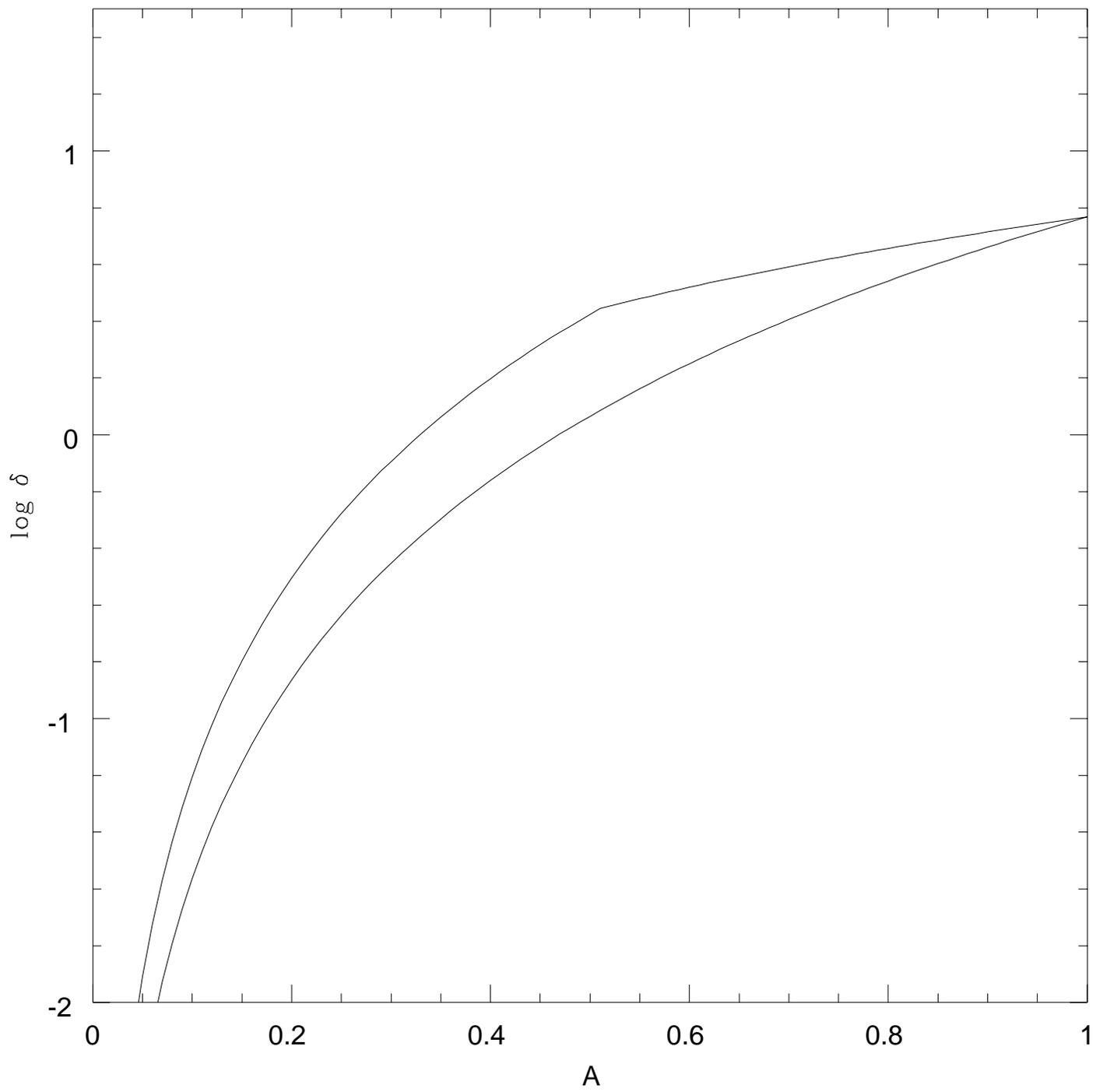
\end{document}